%% file: english_main.tex
\documentclass[12pt]{article}
\usepackage{authblk}
\usepackage[top=25truemm,bottom=28truemm,left=24truemm,right=24truemm]{geometry}
\usepackage{setspace}
\normalsize
\usepackage{amsmath,amssymb,mathrsfs,amsbsy,latexsym,amsfonts,amsthm,bm,fancyhdr,fancybox,titlesec,ulem,comment,appendix}
\usepackage{physics}
\usepackage{color}
\usepackage[dvipdfmx,colorlinks=true,linkcolor=blue,citecolor=cyan]{hyperref}
\usepackage{physics}

\usepackage{graphicx}
\usepackage{color}
\usepackage{comment}
\usepackage{here}
\usepackage{braket}
\usepackage[utf8]{inputenc}
\usepackage{mathtools}

\newcommand{\nn}{\nonumber\\}

\usepackage{epsf}

\numberwithin{equation}{section}
\newcommand{\NT}[1]{{#1}}
\newcommand{\MD}[1]{{#1}}
\newcommand{\PN}[1]{{#1}}
\newcommand{\NTmod}[1]{{#1}}

\title{\bf 
Holographic Description of Bulk Wave Packets in AdS\(_4\)/CFT\(_3\)
}

\author[1]{Norihiro~Tanahashi\thanks{\tt tanahashi(at)gauge.scphys.kyoto-u.ac.jp}}
\author[2]{Seiji~Terashima\thanks{\tt terasima(at)yukawa.kyoto-u.ac.jp}}
\author[1]{Shiki~Yoshikawa\thanks{\tt yoshikawa.shiki(at)gauge.scphys.kyoto-u.ac.jp}}
\vspace{5mm}
\affil[1]{\it\normalsize Department of Physics, Kyoto University, Kyoto 606-8502, Japan}
\affil[2]{\it\normalsize 
Center for Gravitational Physics and Quantum Information,  
\mbox{Yukawa Institute for Theoretical Physics, Kyoto University, Kyoto 606-8502, Japan}  }
\date{\today}

\begin{document}

\maketitle
\thispagestyle{fancy}
\renewcommand{\headrulewidth}{0pt}

\begin{abstract}
In this paper, we extend the study of wave packets from the AdS$_3$/CFT$_2$ correspondence to AdS$_4$/CFT$_3$ and examine properties of their energy density. We find that, while the energy still localizes on the light cone, its spatial distribution exhibits momentum dependence and is no longer localized in higher dimensions. This result is significant as it reflects leading $1/N$ corrections to the generalized free field theory associated with the free bulk theory at \(N=\infty\). Our findings are consistent with previous studies on entanglement wedge reconstruction including $1/N$ corrections, and also provide new insights into the structure of wave packets in higher-dimensional holography.
\end{abstract}
\newpage
\thispagestyle{empty}
\setcounter{tocdepth}{2}

\setlength{\abovedisplayskip}{12pt}
\setlength{\belowdisplayskip}{12pt}

\tableofcontents
\newpage

\pagestyle{plain}

\section{Introduction and summary}

The AdS/CFT correspondence~\cite{Maldacena:1997re} is one of the most promising approaches to studying quantum gravity. 
In particular,
according to this correspondence, we have concrete examples of quantum gravities in terms of CFT.
%
This
correspondence is commonly formulated through the GKPW relation~\cite{Gubser:1998bc,Witten:1998qj},
which equates the CFT partition function with sources to the AdS quantum gravity partition function with corresponding boundary conditions. This is particularly useful in Euclidean AdS.

Another approach, based on the operator formalism, identifies the Hilbert spaces and Hamiltonians of both theories 
\NT{with a given foliation of spacetime.}
This was initiated in~\cite{Balasubramanian:1998sn,Banks:1998dd}.
In~\cite{Banks:1998dd}, it was noted that the large $N$ factorized energy spectrum of CFT matches the free theory spectrum in AdS, as both represent conformal symmetry.
Such an approach to understanding AdS/CFT 
in the operator formalism was clarified and further developed in~\cite{Terashima:2017gmc,Terashima:2019wed}.
This understanding in the operator formalism is intuitive as it directly deals with states and is likely important for understanding quantum gravity.


In such studies, wave packets in bulk spacetime are
particularly
important
because 
wave packets serve as fundamental probes of the bulk and play a crucial role in understanding the relationship between the bulk spacetime and the CFT, 
\NT{which would be useful to investigate properties of}
the quantum gravitational spacetime.
\NT{Indeed, the bulk wave packet in AdS/CFT has been studied in~\cite{Terashima:2020uqu,Terashima:2021klf} with respect to the behavior on the CFT side,}
and further examined in~\cite{Terashima:2023mcr,Kinoshita:2023hgc}%
.\footnote{
In~\cite{Horowitz:1999gf}, the black holes moving in AdS were considered and some of those correspond to wave packets.
In~\cite{Nozaki:2013wia}, they are realized by the local quench of a heavy primary operator in CFT.}
However, \NT{properties of}
the wave packets remain relatively unexplored despite their importance. 

In a previous study~\cite{Terashima:2023mcr}, wave packet states were constructed in the AdS/CFT correspondence, and the energy density was explicitly calculated in the simplest case of AdS$_3$/CFT$_2$. 
It was shown that the energy density on the CFT side localizes on the light cone, which can be interpreted as 
two ``particle-like'' objects propagating at the light speed in the CFT. 
Calculating the energy density, which is one of the most fundamental non-vanishing three-point functions in CFT, is particularly significant as it corresponds to including the leading $1/N$ corrections to the generalized free field theory (GFF)~\cite{Duetsch:2002hc}
associated with the free bulk theory at \(N=\infty\).


\MD{
In this study, we extend the analysis to a higher-dimensional setting, specifically to the AdS$_4$/CFT$_3$ correspondence. We construct wave packets propagating in AdS$_4$ and, using the AdS/CFT dictionary, compute the energy density in the dual CFT$_3$. In this setup, we clarify the distribution of energy density and examine its differences from the AdS$_3$/CFT$_2$ case. We find that the energy density in CFT$_3$ remains localized on the light cone. Moreover, we analyze the momentum dependence of the energy distribution \NT{on this light cone} and find that, in contrast to the AdS\(_3\)/CFT\(_2\) case, the energy is no longer spatially localized in higher dimensions.  
}

\MD{
In the previous study on AdS$_3$/CFT$_2$, a particle-like description
\NT{naturally emerged on the CFT side as a counterpart of a bulk wave packet.}
However, in the case of AdS$_4$/CFT$_3$, the energy is spatially spread at a given time, and then a particle-like description does not 
\NT{apply.}
Additionally, our results are consistent with a weaker version of entanglement wedge reconstruction~\cite{Cotler:2017erl, Terashima:2023mcr, Sugishita:2023wjm, Sugishita:2024lee}.  
Our finding provides a concrete example \NT{of time evolution of wave packet states and their behavior at $O(1/N)$ order} in higher dimensions, \NT{where bulk gravity becomes dynamical}. Notably, the results in this paper rely solely on the BDHM extrapolation relation~\cite{Banks:1998dd} and are derived analytically using the known three-point function in three-dimensional CFT~\cite{Osborn:1993cr}.
}

\MD{This paper is organized as follows. In Section~\ref{sec:wave-packets}, we construct the wave packet states in the AdS/CFT correspondence. In Section~\ref{sec:energy-density}, we calculate the energy density. The detailed computations for this section are provided in the Appendix~\ref{cal}. Finally, Section~\ref{sec:discussion} discusses the implications of our results.}

\section{Wave packets in AdS/CFT}
\label{sec:wave-packets}

We construct wave packet states in the AdS/CFT correspondence following~\cite{Terashima:2023mcr}. The AdS/CFT correspondence implies that 
the bulk Hilbert space in AdS 
corresponds to
the Hilbert space of the dual conformal field theory (CFT). 
Then, we prepare a single-particle state in the bulk theory, which 
\NT{is represented by}
a bulk wave packet and is equivalently described as a CFT state.
Specifically, we will focus on 
Gaussian wave packets in the following analysis.

In \((d+1)\)-dimensional Minkowski spacetime, the wave packet for a free scalar field $\phi$ at \(t = \Vec{x} = 0\) can be expressed as
\begin{align}
    \int d\Vec{x} \, e^{-\frac{\Vec{x}^2}{2a^2} + i\Vec{p} \cdot \Vec{x}} \phi(t = 0, \Vec{x}) \ket{0},
    \label{eq:minkowski_wavepacket}
\end{align}
where \(\Vec{p}\) represents the momentum of the wave packet.
Instead of specifying the momentum of a space-like direction, say $x_1$, 
we can specify the energy because they are related by the on-shell condition \NT{$\omega^2 - \vec p\,{}^2 = 0$}. 
Then, the wave packet
integrated over time and  spatial directions \(x^i\) (\(i = 2, \ldots, d\)), is given by
\begin{align}
    \int dt \prod_{i = 2}^{d} dx^i \, e^{-\frac{x^i x_i + t^2}{2a^2} + i p_i x^i + i \omega t} \phi(t, \Vec{x})\big|_{x_1 = 0} \ket{0},
    \label{eq:ads_wavepacket}
\end{align}
where \(i\) runs over directions except for 
\(x_1\)
and the size of wave packet is given by \(a\). Here, we assume \(a \,|p^i| \gg 1\) and \(a \,\omega \gg 1\)
to ensure
that the wave packet has a definite 
\NT{propagation direction}
with the momentum \(p_i\) and energy \(\omega\).

Based on the above construction, let us now consider wave packets in the AdS/CFT correspondence,
using
a scale field \(\phi\) in the Poincare patch of the \((d+1)\)-dimensional AdS spacetime.
In the AdS spacetime, the wave packet can be well-approximated by the Minkowski wave packet if the size of wave packet \(a\) is much smaller than the AdS radius \(l_{\mathrm{AdS}} = 1\), i.e., \(a \ll 1\). 
At the boundary, the bulk scalar field \(\phi\) is related to the CFT primary operator \(\mathcal{O}\) via the BDHM relation~\cite{Banks:1998dd}.
Thus we can write the bulk wave packet state at the boundary as  
\begin{align}
    \ket{p, \omega} 
    &= \lim_{z \to 0} \frac{1}{z^\Delta} 
    \int dt \, d^{d-1} x \, e^{-\frac{x^i x_i + t^2}{2a^2} + i p_i x^i - i \omega t} \phi(t, z, x^i) \ket{0} \nonumber \\
    &= \int dt \, d^{d-1} x \, e^{-\frac{x^i x_i + t^2}{2a^2} + i p_i x^i - i \omega t} \mathcal{O}(t, x) \ket{0},
    \label{wads}
\end{align}
where we have identified \(x^1 = z\), and \(z \to 0\) corresponds to the AdS boundary in the Poincaré AdS coordinates. Here, \(\mathcal{O}(t, x)\) is the boundary CFT operator dual to the bulk scalar field \(\phi(t, z, x^i)\).
The wave packet inside the bulk is given by the time evolution of this state \(\ket{p, \omega}\), and 
the time evolution of the bulk wave packet state follows a light-like trajectory.
The size of the wave packet in spacetime remains \(\mathcal{O}(a)\) during this evolution.

We note that this state is not a wave packet state in the CFT,
even for the free CFT, because
both the energy and the momenta are fixed and the on-shell condition is not satisfied generically.
Note also that this state is well-defined because of the smearing in the time direction,
not just the spacial directions~\cite{Nagano:2021tbu}.

\section{Energy density of wave packets in CFT picture}
\label{sec:energy-density}

The time evolution of the bulk wave packet state~\eqref{wads} can be computed in the dual CFT. Bulk wave packets serve as fundamental probes of bulk spacetime, making them crucial for understanding how the bulk geometry emerges from the CFT. It is therefore important to investigate the CFT counterpart of the bulk wave packet's behavior.

Here, we study the distribution of the wave packet state~\eqref{wads} in the CFT by calculating the time evolution of the energy density.\footnote{
Other observables, such as the expectation value of the primary scalar operator \(\mathcal{O}\), could also be considered. 
However, the three-point functions of the scalar primaries depend on the theory
and not universal.
For the non-zero three-point function case, the result may be similar to the energy density.
Instead of such quantities, 
we can consider the overlap between the wave packet state and \( {\cal O} (t,x)\ket{0}\).
However, 
this vanishes for $t\neq 0$
and are therefore not suitable for the study of where the CFT state localized. 
For a detailed discussion, see~\cite{Terashima:2023mcr}.
}
The energy density of the wave packet state~\eqref{wads} is determined by the three-point function involving two primary operators and the energy-momentum tensor:
\begin{align}
    & \bra{p, \omega} T_{00}(t = \bar{t}, x^i = \bar{x}^i) \ket{p, \omega} \nonumber \\
    =& \int dt_1 \, d^d x_1 \, e^{-\frac{(x_1^i)^2 + t_1^2}{2a^2} - i p_i x_1^i + i \omega t_1} 
    \int dt_2 \, d^d x_2 \, e^{-\frac{(x_2^i)^2 + t_2^2}{2a^2} + i p_i x_2^i - i \omega t_2} \nonumber \\
    &\quad \times \bra{0} \mathcal{O}(t_1, x_1) T_{00}(t = \bar{t}, x^i = \bar{x}^i) \mathcal{O}(t_2, x_2) \ket{0}.
    \label{eq:energy_density}
\end{align}
In the CFT, such three-point functions are well-known and can be computed explicitly.\footnote{
The operator ordering in these three-point functions is not time-ordered. 
We can
compute these \MD{three-point functions} by analytic continuation from the
Euclidean correlators. This can be implemented by adding an infinitesimal imaginary time shift to the insertion times of the operators. Specifically, the prescription is:
\begin{align}
    t_{ij} \rightarrow t_{ij} - i\epsilon_{ij}, \quad \epsilon_{ij} > 0 \; \mathrm{for} \; i < j,
\end{align}
as discussed in~\cite{duffin_lecture}.
}

This calculation does not rely on the generalized free field approximation. It corresponds to the leading order in the large \(N\) expansion and remains valid for sufficiently large but finite \(N\). This distinction is significant as it captures contributions of finite \(N\) effects beyond the generalized free approximation \NT{corresponding to the $N\to\infty$ limit}.

\subsection{Brief review of \texorpdfstring{AdS$_3/$CFT$_2$}{AdS_3/CFT_2} case}
We provide a brief review of the calculations in AdS$_3$/CFT$_2$ presented in~\cite{Terashima:2023mcr}. What we want to calculate is given by 
\begin{align}
   & \bra{p,\omega} T_{00}(t=\bar{t}, x=\bar{x}) \ket{p,\omega} \nn
   =
   & \int d t_1 \, d x_1 \, e^{-\frac{(x_1)^2+t_1^2}{2 a^2} - i p x_1 + i \omega t_1} 
   \int d t_2 \, d x_2 \, e^{-\frac{(x_2)^2+t_2^2}{2 a^2} + i p x_2 - i \omega t_2} \nn
   & \times \bra{0} {\cal O}(t_1,x_1) T_{00}(t=\bar{t}, x=\bar{x}) {\cal O}(t_2,x_2) \ket{0}.
\end{align}

To evaluate the three-point function
\(\bra{0} {\cal O}(t_1,x_1) T_{00}(t=\bar{t}, x=\bar{x}) {\cal O}(t_2,x_2) \ket{0}\), 
we analyze it within the Euclidean CFT$_2$ framework. In terms of complex coordinates, the three point function can be written as 
\(\bra{0} {\cal O}(z_1,\bar{z}_1) T_{00}(\xi, \bar{\xi}) {\cal O}(z_2,\bar{z}_2) \ket{0}\), 
where the energy density is expressed as 
\(\frac{1}{2\pi} T_{00}(z,\bar{z}) = \frac{1}{2\pi} (T(z) + \bar{T}(\bar{z}))\).

Using the conformal Ward identity, we find that 
\begin{align}
    \bra{0} T(\xi) {\cal O}_1(z_1,\bar{z}_1) {\cal O}_2(z_2,\bar{z}_2) \ket{0}
    &=
    \sum_{i=1,2} 
    \left( \frac{h_i}{(\xi - z_i)^2} + \frac{\partial_i}{\xi - z_i} \right)
    \bra{0} {\cal O}_1(z_1,\bar{z}_1) {\cal O}_2(z_2,\bar{z}_2) \ket{0} \nn
    &=
    \sum_{i=1,2} 
    \left( \frac{h_i}{(\xi - z_i)^2} + \frac{\partial_i}{\xi - z_i} \right)
    \frac{1}{(z_1 - z_2)^{h_1+h_2} (\bar{z}_1 - \bar{z}_2)^{\bar{h}_1 + \bar{h}_2}},
    \label{ward_identity}
\end{align}
where \(h_i\) and \(\bar{h}_i\) are the holomorphic and antiholomorphic conformal weights of the operators \({\cal O}_i\), and \(\partial_i\) denotes the derivative with respect to \(z_i\). The structure of the correlator is entirely determined by conformal symmetry and the operator dimensions.

Here, the normalization of the two-point function is chosen to be 1, and \(h\) represents the weight of the primary operator. The conformal dimension is given by \(\Delta = h + \bar{h}\), and the spin 
\NT{$s = h - \bar h$}
of the \NT{scalar} primary operator \(\mathcal{O}\) is \(0\).
Continuing to Minkowski signature, we will replace \(z \rightarrow u = t + x\) and \(\bar{z} \rightarrow -v = -(t - x)\),\footnote{\NT{This definition of the coordinates $u,v$ are opposite from the standard definition $u=t-x, v = t+x$. We keep using our definition of $u,v$ to facilitate comparison with results of \cite{Terashima:2023mcr}.}}
then we obtain the following expression:
\begin{align}
    & \bra{0} {\cal O}(t_1,x_1) T_{00}(t=\bar{t}, x=\bar{x})
   {\cal O}(t_2,x_2) \ket{0} \nn
   = &
    \bra{0}  {\cal O}(u_1, v_1) 
    ( T (\bar{u}) + \bar{T} (\bar{v}))
    {\cal O}(u_2, v_2) \ket{0} \nn
    = & 
      \sum_{i=1,2} 
    \left( \frac{\Delta/2}{(\bar{u}-u_i)^2} + \frac{\partial_{u_i}}{\bar{u}-u_i} + 
    \frac{\Delta/2}{(\bar{v}-v_i)^2} + \frac{\partial_{v_i}}{\bar{v}-v_i} \right)
    \frac{1}{(u_1-u_2)^{\Delta} (v_2-v_1)^{\Delta}} \nn
    = & 
      \frac{\Delta}{2} 
     \left(
      \frac{(u_1-u_2)^2}{(\bar{u}-u_1)^2(\bar{u}-u_2)^2}
      +
        \frac{(v_1-v_2)^2}{(\bar{v}-v_1)^2(\bar{v}-v_2)^2}
      \right)
    \frac{1}{(u_1-u_2)^{\Delta} (v_2-v_1)^{\Delta}}.
\end{align}
Using this, the energy density for the bulk wave packet state 
becomes
\begin{align}
   & \bra{p,\omega}   T_{00}(t=\bar{t}, x=\bar{x}) \ket{p,\omega} \nn
   =
   & \int d t_1 \, d x_1 \, e^{-\frac{(x_1)^2+t_1^2}{2 a^2} - i p x_1 + i \omega t_1} 
   \int d t_2 \, d x_2 \, e^{-\frac{(x_2)^2+t_2^2}{2 a^2} + i p x_2 - i \omega t_2} \nn
   & \times \frac{\Delta}{2} 
     \left(
      \frac{(u_1-u_2)^2}{(\bar{u}-u_1)^2(\bar{u}-u_2)^2}
      +
        \frac{(v_1-v_2)^2}{(\bar{v}-v_1)^2(\bar{v}-v_2)^2}
      \right)
    \frac{1}{(u_1-u_2)^{\Delta} (v_2-v_1)^{\Delta}}.
    \label{T00-orig_AdS3CFT2}
\end{align}

In the following, we will compute the energy density explicitly.
Before showing the detailed derivations,
we present the final result; the energy density \({\cal E}(\bar{t},\bar{x})\) is given by
\begin{align}
   {\cal E}(\bar{t},\bar{x})
   \coloneqq & \frac{\bra{p,\omega}\MD{\frac{1}{2\pi}} T_{00}(t=\bar{t}, x=\bar{x}) \ket{p,\omega}}{\bra{p,\omega}   p,\omega \rangle} \nn
      \simeq & 
  \frac{1}{2 \sqrt{2 \pi} a}
  \left( e^{-\frac{ (\bar{x}+\bar{t})^2}{2 a^2}} (\omega - p) + e^{-\frac{ (\bar{x}-\bar{t})^2}{2 a^2}} (\omega + p) \right).
\end{align}
This result indicates that the energy is localized on the light cone\,\((\bar{x}=\bar{t}, \bar{x}=-\bar{t})\).

Below, we will demonstrate the calculation based on~\cite{Terashima:2023mcr}. The energy density \eqref{T00-orig_AdS3CFT2} is computed as
\begin{align}
   & \bra{p,\omega}   T_{00}(t=\bar{t}, x=\bar{x}) \ket{p,\omega} \nn
   =
   & \int d t_1 \, d x_1 \, e^{-\frac{ (x_1)^2+t_1^2}{2 a^2}-i p x_1+i {\omega} t_1} 
   \int d t_2 \, d x_2 \, e^{-\frac{ (x_2)^2+t_2^2}{2 a^2}+i p x_2-i {\omega} t_2}  \nn
   & \times \frac{\Delta}{2} 
     \left(
      \frac{(u_1-u_2)^2}{(\bar{u}-u_1)^2(\bar{u}-u_2)^2}
      +
        \frac{(v_1-v_2)^2}{(\bar{v}-v_1)^2(\bar{v}-v_2)^2}
      \right)
    \frac{1}{(u_1-u_2)^{ \Delta} (v_2-v_1)^{ \Delta} } \nn
    =
    & \frac{1}{4}\int d u_1 \, d v_1 \, d u_2 \, d v_2 \,\, e^{-\frac{ (u_1)^2+(v_1)^2+(u_2)^2+(v_2)^2}{4 a^2}+i (p_u u_1+p_v v_1-p_u u_2-p_v v_2)/2}
     \nn
   & \times \frac{\Delta}{2} 
     \left(
      \frac{(u_1-u_2)^2}{(\bar{u}-u_1)^2(\bar{u}-u_2)^2}
      +
        \frac{(v_1-v_2)^2}{(\bar{v}-v_1)^2(\bar{v}-v_2)^2}
      \right)
    \frac{1}{(u_1-u_2)^{ \Delta} (v_2-v_1)^{ \Delta} } \nn
    =
    & \frac{1}{4}\int d u_1 \, d v_1 \, d u_2 \, d v_2 \,\, e^{-\frac{ (u_1-i p_u a^2)^2+(v_1-i p_v a^2)^2+(u_2+i p_u a^2)^2+(v_2+i p_v a^2)^2}{4 a^2} 
    - \frac{a^2}{2} ((p_u)^2+(p_v)^2) }
     \nn
   & \times \frac{\Delta}{2} 
     \left(
      \frac{(u_1-u_2)^2}{(\bar{u}-u_1)^2(\bar{u}-u_2)^2}
      +
        \frac{(v_1-v_2)^2}{(\bar{v}-v_1)^2(\bar{v}-v_2)^2}
      \right)
    \frac{1}{(u_1-u_2)^{ \Delta} (v_2-v_1)^{ \Delta} },
    \label{T00_AdS3CFT2}
\end{align}
where we defined \(p_u \coloneqq \omega - p \,\, (\gg 0)\) and \(p_v \coloneqq \omega + p \,\, (\gg 0)\). 

We evaluate the part of the expression corresponding to the first term in the parentheses as
\begin{align}
    A \coloneqq & \int d u_1 \, d v_1 \, d u_2 \, d v_2 \,\, e^{-\frac{ (u_1-i p_u a^2)^2+(v_1-i p_v a^2)^2+(u_2+i p_u a^2)^2+(v_2+i p_v a^2)^2}{4 a^2} 
    - \frac{a^2}{2} \big((p_u)^2+(p_v)^2\big) }
     \nn
   & \times  
      \frac{(u_1-u_2)^2}{(\bar{u}-u_1)^2(\bar{u}-u_2)^2}
    \frac{1}{(u_1-u_2)^{ \Delta} (v_2-v_1)^{ \Delta} } \nn
    = & 
    \,e^{-\frac{a^2}{2} (p_v)^2 } 
    \int  d v_1  \, d v_2 \,\, e^{-\frac{ (v_1-i p_v a^2)^2+(v_2+i p_v a^2)^2}{4 a^2} 
    } 
    \frac{1}{(v_2-v_1)^{ \Delta}}
    \nn
    & \times
    e^{-\frac{a^2}{2} (p_u)^2 } 
    \int d u_1 \,  d u_2  \,\, e^{-\frac{ (u_1-i p_u a^2)^2+(u_2+i p_u a^2)^2}{4 a^2}  }
    \frac{1}{(\bar{u}-u_1)^2(\bar{u}-u_2)^2}
    \frac{1}{(u_1-u_2)^{ \Delta-2}  }.
    \label{A_AdS3CFT2}
\end{align}
The 
other
parts of the calculation can be obtained in a parallel manner. When analytically continuing from Euclidean to Minkowski spacetime, the operator ordering should be fixed. This is implemented using the \(i \epsilon\)-prescription, where the time coordinates are shifted as \(t_i \rightarrow t_i - i \epsilon_i\), with \(\epsilon_1 = -\epsilon_2 = \epsilon > 0\). 

First, let us evaluate the integral over \(v_1\) of \eqref{A_AdS3CFT2}. To do that, we rewrite it as 
\begin{align}
   & e^{-\frac{a^2}{2} ((p_v)^2) } 
    \int  d v_1  \, d v_2 \,\, e^{-\frac{ (v_1-i p_v a^2)^2+(v_2+i p_v a^2)^2}{4 a^2} 
    } 
    \frac{1}{(v_2-v_1)^{ \Delta}} \nn
    =& (-1)^{\Delta} e^{-\frac{a^2}{2} ((p_v)^2) } 
    \int  d v_1  \, d v_2 \,\,  
    \frac{1}{(v_1-v_2)^{ \Delta-q}}
    \frac{\Gamma(\Delta-q)}{\Gamma(\Delta)}
    \frac{\partial^q}{\partial v_1^q} \, e^{-\frac{ (v_1-i p_v a^2)^2+(v_2+i p_v a^2)^2}{4 a^2} 
    },
    \label{no1}
\end{align}
where \(q\) is an integer satisfying \(1 \geq \Delta - q > 0\), and the integration by parts has been performed \(q\) times.

\begin{figure}
    \centering
    \includegraphics[width=0.5\linewidth]{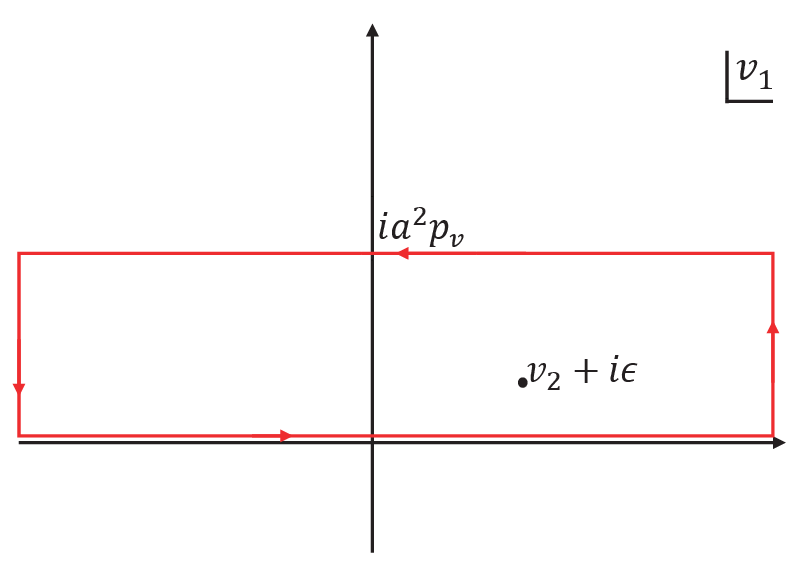}
    \caption{Contour of the \(v_1\) integral.}
    \label{f10}
\end{figure}

By shifting the integration contour for \(v_1\) to \(v_1 \in \mathbb{R} + i a^2 p_v\), the 
\NT{exponential factor in the}
integrand becomes independent of \(p_v\). Since the Gaussian factor \(e^{-\frac{a^2}{2} (p_v)^2}\) suppresses the contribution along this path, it can be neglected compared with the contribution at the pole. Thus, the poles at \(v_1 = v_2\) can be evaluated using the residue theorem by enclosing them with the shifted contour (ensuring residue is within the closed path due to the \(i\epsilon\)-prescription). The contour is illustrated in Figure~\ref{f10}.
If \(\Delta - q\) is not an integer, branch cuts arise, but the integral can still be determined by analytic continuation from the integer case~\cite{Terashima:2023mcr}.

Setting \(q = \Delta - 1\), 
 the integral can be evaluated using the residue theorem as
\begin{align}
  & (-1)^{\Delta} \frac{2 \pi i }{\Gamma(\Delta)}
  e^{-\frac{a^2}{2} (p_v)^2 } 
    \int d v_2 \,\,  
    \left.
    \frac{\partial^{\Delta-1}}{\partial v_1^{\Delta-1}} \, e^{-\frac{ (v_1-i p_v a^2)^2+(v_2+i p_v a^2)^2}{4 a^2} 
    } \right|_{v_1 =v_2} \nn
    \simeq & 
    \NT{(-1)^{\Delta}}
    \frac{2 \pi i }{\Gamma(\Delta)}
  ( -i p_v/2)^{\Delta-1} \,
  e^{-\frac{a^2}{2} (p_v)^2 } 
    \int d v_2 \,\,  
     e^{-\frac{ (v_2)^2+(p_v a^2)^2}{2 a^2} 
    } \nn
    = & 
    \frac{(2 \pi)^{3/2}  }{\Gamma(\Delta)} (-i)^{\Delta}
  a ( p_v/2)^{\Delta-1} \,
    ,
    \label{di}
\end{align}
where only the leading contributions have been retained,
\NT{using the fact that we can approximately evaluate \(|v_2| \leq \mathcal{O}(a)\) because of the Gaussian factor, and then $v_2$ becomes subdominant to $p_v$.}

Next, we compute the following integral:
\begin{align}
  & e^{-\frac{a^2}{2} (p_u)^2 } 
    \int d u_1 \,  d u_2  \,\, e^{-\frac{ (u_1-i p_u a^2)^2+(u_2+i p_u a^2)^2}{4 a^2}  }
    \frac{1}{(\bar{u}-u_1)^2(\bar{u}-u_2)^2}
    \frac{1}{(u_1-u_2)^{ \Delta-2}  } \nn
    =& e^{-\frac{a^2}{2} (p_u)^2 } 
    \int  d u_2  \,\, e^{-\frac{ (u_2+i p_u a^2)^2}{4 a^2}  }
    \frac{1}{(\bar{u}-u_2)^2}
    \int d u_1 \,
    \frac{1}{(u_1-\bar{u})}
    \frac{\partial}{\partial u_1} \left(
     e^{-\frac{ (u_1-i p_u a^2)^2}{4 a^2}  }
    \frac{1}{(u_1-u_2)^{ \Delta-2}  } \right).
    \label{u1u2}
\end{align} 
Similar to the \(v_1\) integral, the \(u_1\) integral is evaluated by shifting the integration path to \(u_1 \in \mathbb{R} + i a^2 p_u\)
and applying the residue theorem to the pole at $u_1 = \bar u$.\footnote{
The pole at \(u_1 = u_2\) can be ignored as it leaves only a small Gaussian factor \(e^{-\frac{a^2}{2} (p_u)^2}\).
}
The result of the integral is given by
\begin{align}
 & 2 \pi i e^{-\frac{a^2}{2} (p_u)^2 } e^{-\frac{ (\bar{u}-i p_u a^2)^2}{4 a^2}  }
    \int  d u_2  \,\, e^{-\frac{ (u_2+i p_u a^2)^2}{4 a^2}  }
    \frac{1}{(\bar{u}-u_2)^\Delta}
     \left(  i p_u/2-(\Delta-2) 
     \frac{1}{(\bar{u}-u_2)}
    \right),
\end{align}
where we have ignored terms proportional to \(\bar{u}\) since, as will be shown later, the Gaussian factor \(e^{-\frac{ (\bar{u})^2}{2 a^2}  }\) suppresses contributions from large \(\bar{u}\).
For the \(u_2\) integral, the path is shifted to \(u_2 \in \mathbb{R} - i a^2 p_u\). Using this path, the residue at \(u_2 = \bar{u}\) can be evaluated in the same way. 

 From the above calculations, the result for \(A\) is approximately given by 
\begin{align} 
    A \simeq \, & e^{-\frac{ (\bar{u})^2}{2 a^2}  }
      (2 \pi)^{5/2}  
      \frac{1}{\Gamma(\Delta)^2}
      (p_v p_u/4)^{\Delta-1}
      \frac{p_u a}{\Delta}.
\end{align}
\NT{The contribution from the second term in the parentheses of \eqref{T00_AdS3CFT2} is given by the above expression with $(\bar u, p_u)$ and $(\bar v, p_v)$ are swapped with each other.}

The normalization factor can also be computed similarly:
\begin{align}
    {\cal N}^2 = &\bra{p,\bar\omega}   p,\bar\omega \rangle \nn
   =
   & \int d t_1 \, d x_1 \, e^{-\frac{ (x_1)^2+t_1^2}{2 a^2}-i p x_1+i \bar{\omega} t_1} 
   \int d t_2 \, d x_2 \, e^{-\frac{ (x_2)^2+t_2^2}{2 a^2}+i p x_2-i \bar{\omega} t_2} 
    \frac{1}{(u_1-u_2)^{ \Delta} (v_2-v_1)^{ \Delta} } \nn
    =
    & \frac{1}{4}\int d u_1 \, d v_1 \, d u_2 \, d v_2 \,\, e^{-\frac{ (u_1-i p_u a^2)^2+(v_1-i p_v a^2)^2+(u_2+i p_u a^2)^2+(v_2+i p_v a^2)^2}{4 a^2} 
    - \frac{a^2}{2} ((p_u)^2+(p_v)^2) }
     \nn
   & \times 
    \frac{1}{(u_1-u_2)^{ \Delta} (v_2-v_1)^{ \Delta}} \nn
    \simeq
    & \frac{1}{4} \,\,
     \frac{(2 \pi)^{3/2}  }{\Gamma(\Delta)} (-i)^{\Delta}
      a (p_v/2)^{\Delta-1} \times 
      \frac{(2 \pi)^{3/2}  }{\Gamma(\Delta)} (i)^{\Delta}
      a (p_u/2)^{\Delta-1}.
\end{align}

Using these results, the energy density can be expressed as:
\begin{align}
   {\cal E}(\bar{t},\bar{x}) = &  \frac{1}{{\cal N}^2} \, \bra{p,\omega} \frac{1}{2 \pi}  T_{00}(t=\bar{t}, x=\bar{x}) \ket{p,\omega} \nn
      \simeq & 
  \frac{1}{2 \sqrt{2 \pi} a}
  \left( e^{-\frac{ (\bar{u})^2}{2 a^2}  } p_u + e^{-\frac{ (\bar{v})^2}{2 a^2}  } p_v \right).
  \label{calE_AdS3CFT2}
\end{align}
The energy in the vicinity of \(\bar{u}=0\) and \(\bar{v}=0\) is found to be \(p_u/2\) and \(p_v/2\), respectively. Adding these contributions gives the total energy \(\omega\), which is consistent with the wave packet state under consideration.

\subsection{\texorpdfstring{AdS$_4/$CFT$_3$}{AdS_4/CFT_3} case}
\label{sec:AdS4CFT3}

As a generalization of the analysis of AdS\(_3\)/CFT\(_2\) given in the previous section, we consider the case of AdS\(_4\)/CFT\(_3\) in this section. 
In this case, the increase in spatial dimensions introduces a nontrivial angular distribution of energy. In AdS\(_3\)/CFT\(_2\), where the spatial direction is one-dimensional,
\NT{the fact that the energy density localizes on the light cone implies that it is spatially localized.}
However, in AdS\(_4\)/CFT\(_3\), even if localization on the light cone is achieved, it is not immediately clear whether the energy remains spatially localized.

\begin{figure}[htbp]
  \begin{minipage}[t]{0.45\linewidth}
    \centering
    \includegraphics[width=4.5cm]{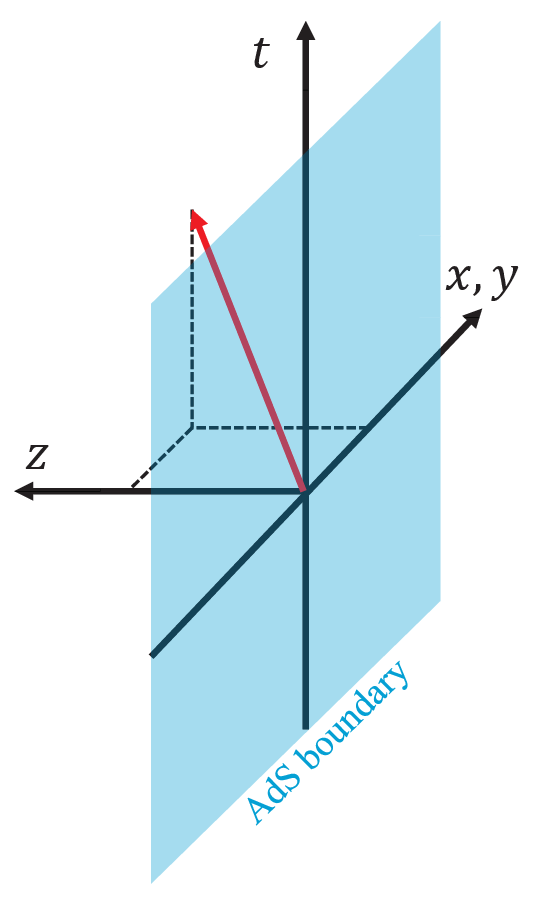}
    \caption{A wave packet propagating from the boundary into the bulk. The red arrow indicates the wave packet trajectory.}
    \label{f5}
  \end{minipage}
\hspace{10mm} 
  \begin{minipage}[t]{0.45\linewidth}
    \centering
    \includegraphics[width=6.5cm]{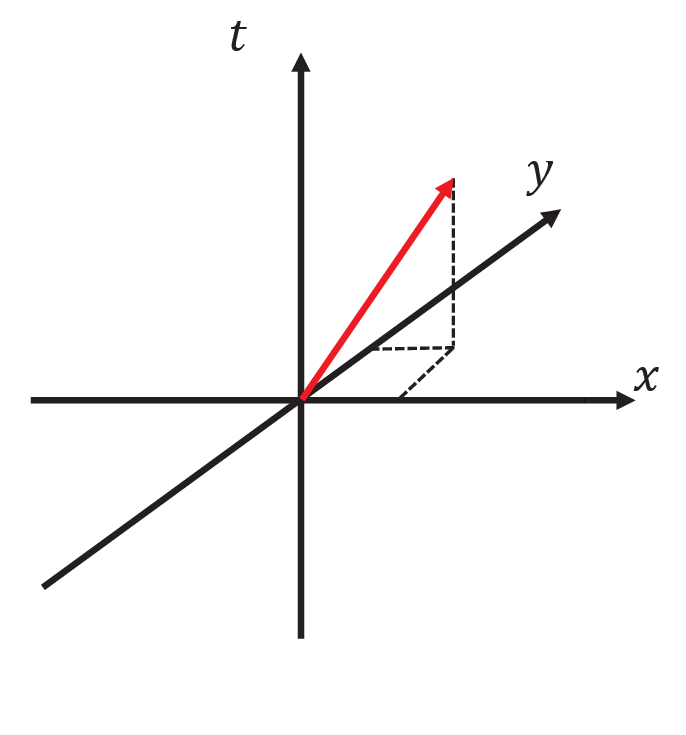}
    \caption{The corresponding setup on the CFT side. The red arrow
    indicates \( \omega, p_x, p_y\),
    while
    the state is neither a wave packet nor a particle in CFT.}
    \label{f6}
  \end{minipage}
\end{figure}

The wave packet state can be expressed as follows~\cite{Terashima:2023mcr}:
\begin{align}
    \ket{\omega, p_x, p_y}
    &= \lim_{z \rightarrow 0} {1 \over z^\Delta}
    \int d t \, d x \, d y \, e^{-\frac{t^2 + x^2 + y^2}{2 a^2} - i \omega t + i p_x x
    \NT{+ i p_y y}
    } 
    \phi(t, x, y, z) \ket{0} \nn
    &= \int d t \, d x \, d y \, e^{-\frac{t^2 + x^2 + y^2}{2 a^2} - i \omega t + i p_x x + i p_y y} {\cal O}(t, x, y) \ket{0},
    \label{wads3}
\end{align}
where \(z\) is the bulk radial coordinate, and the limit \(z \to 0\) corresponds to the asymptotic boundary of the AdS spacetime.
We assume
\begin{align}
    a^2 \omega^2 \gg 1, \quad a^2 p_x^2 \gg 1, \quad a^2 p_y^2 \gg 1. \label{asmp}
\end{align}
Additionally,
we require  \(\omega^2 - p_x^2 - p_y^2 > 0\) because 
the wave packet corresponds to 
\NT{a bulk wave packet}
that satisfies
the on-shell condition \(\omega^2 - p_x^2 - p_y^2 - p_z^2 = 0\).
This setup is illustrated in Figures~\ref{f5} and \ref{f6}. 

To obtain the energy density, we calculate the following quantity:
\begin{align}
   & \bra{\omega, p_x, p_y} T_{00}(t=\bar{t}, x=\bar{x}, y=\bar{y}) \ket{\omega, p_x, p_y} \nn
   =
   & \int d t_1 \, d x_1 \, d y_1 \, e^{-\frac{t_1^2 + x_1^2 + y_1^2}{2 a^2} + i \omega t_1 - i p_x x_1 - i p_y y_1} 
    \int d t_3 \, d x_3 \, d y_3 \, e^{-\frac{t_3^2 + x_3^2 + y_3^2}{2 a^2} - i \omega t_3 + i p_x x_3 + i p_y y_3}  \nn
   & \times \bra{0} {\cal O}(t_1, x_1, y_1) T_{00}(t=\bar{t}, x=\bar{x}, y=\bar{y})
   {\cal O}(t_3, x_3, y_3) \ket{0}.
   \label{T00_AdS4CFT3}
   \end{align}
The result can be summarized as follows:
\begin{align}
     {\cal{E}}(t, x, y; \omega, p_x, p_y)  
    \simeq & \frac{(\omega^2 - p_x^2 - p_y^2)^{\frac{3}{2}}}{\sqrt{2}\pi^\frac{3}{2}a(t+\sqrt{x^2+y^2})\left(\omega - p_x\frac{x}{\sqrt{x^2+y^2}}-p_y\frac{y}{\sqrt{x^2+y^2}}\right)^2}e^{-\frac{(t-\sqrt{x^2+y^2})^2}{2a^2}}.
\end{align}
This expression shows that the energy density is localized on the light cone. While the distribution is maximized along the momentum direction, it is not localized there.

The derivation proceeds as follows.
In a \(d\)-dimensional Euclidean CFT, the following expression is known~\cite{Osborn:1993cr, Belin:2019mnx}:
\begin{align}
    &\braket{T_{\mu\nu}(x_1)O(x_2)O(x_3)} = \frac{C_{TOO}}{x_{12}^d x_{13}^d x_{23}^{2\Delta-d}} t_{\mu\nu}(X), \\
    &t_{\mu\nu}(X) = \frac{X_{\mu} X_{\nu}}{X^2} - \frac{1}{d} \delta_{\mu\nu}, \qquad 
    X_{\mu} = \frac{(x_{12})_{\mu}}{x_{12}^2} - \frac{(x_{13})_{\mu}}{x_{13}^2},
\end{align}
\NT{where \(C_{TOO}\) is the normalization constant.}
Here, \( x_{ij}^\mu \coloneqq x_i^\mu - x_j^\mu \), and 
\( x_{ij}^2 = (t_i - t_j)^2 + (x_i - x_j)^2 + (y_i - y_j)^2 + \cdots \) in the Euclidean case.
Using this expression \NT{for $d=3$}, we find
\begin{align}
   & \bra{0} {\cal O}(t_1, x_1, y_1) T_{00}(t_2, x_2, y_2) {\cal O}(t_3, x_3, y_3) \ket{0} = A + B + C + D,
\end{align}
where the terms \(A, B, C\) and \(D\) are given by
\begin{align}
   & A \coloneqq  \frac{C_{TOO}}{x_{21}^{5} x_{13}^{2 \Delta - 1} x_{23}} t_{21}^2, \\
   & B \coloneqq  -\frac{2 C_{TOO}}{x_{21}^{3} x_{13}^{2 \Delta - 1} x_{23}^{3}} t_{21} t_{23}, \\
   & C \coloneqq  \frac{C_{TOO}}{x_{21} x_{13}^{2 \Delta - 1} x_{23}^{5}} t_{23}^2, \\
   & D \coloneqq  -\frac{C_{TOO}}{3} \frac{1}{x_{21}^{3} x_{13}^{2 \Delta - 3} x_{23}^3}.
\end{align}
When analytically continuing to Minkowski space, the \(i\epsilon\)-prescription for ordering is applied: 
\begin{align}
    t_{ij} \rightarrow t_{ij} - i\epsilon_{ij}, \; \epsilon_{ij} > 0 \quad \mathrm{for} \, i < j.
\end{align}

Since \(A, B, C\) and \(D\) share a common structure, we consider the following general integrand:
\begin{align}
    S(\alpha, \beta, \gamma) \coloneqq \frac{1}{x_{21}^\alpha x_{13}^\gamma x_{23}^\beta}
\NT{
e^{-\frac{t_1^2 + x_1^2 + y_1^2}{2 a^2} + i \omega t_1 - i p_x x_1 - i p_y y_1} 
e^{-\frac{t_3^2 + x_3^2 + y_3^2}{2 a^2} - i \omega t_3 + i p_x x_3 + i p_y y_3}
}.
\end{align}
Below we work in the Lorentzian case, for which  \MD{\( x_{ij}^2 = -(t_i - t_j)^2 + (x_i - x_j)^2 + (y_i - y_j)^2 \).}

\NT{Without loss of generality, we may set}
\(x_2 = (\bar{t}, \bar{x}, 0)\). For the integrand, we use the following approximations:
\begin{align}
    & x_{21}^2 \simeq -(\bar{u} - u_1)(\bar{v} - v_1), \\
    & x_{23}^2 \simeq -(\bar{u} - u_3)(\bar{v} - v_3),
\end{align}
where \(\bar{u} = \bar{t} + \bar{x}, \bar{v} = \bar{t} - \bar{x}, u_i = t_i + x_i, v_i = t_i - x_i\). These approximations are valid under the condition \(\bar{t} \gg a\), which is of interest to us. 
Then, if both \(\bar{u}\) and \(\bar{v}\) are large,
the integrated \( S(\alpha, \beta, \gamma) \) becomes small.
Thus, we will consider 
the case \(\bar{v} \gg a\) and $|\bar{u}| ={\cal O}(a)$.

Due to the Gaussian suppression, 
\NT{$|v_i|=\mathcal{O}(a)$}
and we can further approximate
\begin{align}
    & x_{21}^2 \approx -(\bar{u} - u_1) \bar{v}, \\
    & x_{23}^2 \approx -(\bar{u} - u_3) \bar{v}.
\end{align}

Next, consider the following coordinate transformation:
\begin{align}
    & u = \frac{1}{2}(u_1 - u_3), & \Tilde{u} = \frac{1}{2}(u_1 + u_3), \\
    & v = \frac{1}{2}(v_1 - v_3),  & \Tilde{v} = \frac{1}{2}(v_1 + v_3), \\
    & y = \frac{1}{2}(y_1 - y_3), & \Tilde{y} = \frac{1}{2}(y_1 + y_3).
\end{align}
In this coordinate system, the Gaussian factor becomes
\begin{align}
    e^{-\frac{1}{2a^2}(u^2 + v^2 + \Tilde{u}^2 + \Tilde{v}^2 + 2y^2 + 2\Tilde{y}^2) + i p_u u + i p_v v - 2i p_y y},
\end{align}
where we have defined \(p_u \coloneqq \omega - p_x\) and \(p_v \coloneqq \omega + p_x\).

The integral of \(S\) can be written as
\begin{align}
    \int d^6x \, S = & \MD{(-1)^{\frac{\alpha+\beta+\gamma}{2}}}2\int du\, d\Tilde{u}\, dv\, d\Tilde{v}\, dy\, d\Tilde{y}\, 
    e^{-\frac{1}{2a^2}(u^2 + v^2 + \Tilde{u}^2 + \Tilde{v}^2 + 2y^2 + 2\Tilde{y}^2) + i p_u u + i p_v v - 2i p_y y} \nn
    & \quad \times 
    \frac{1}{(\Tilde{u} + u - \bar{u})^{\frac{\alpha}{2}} \bar{v}^{\frac{\alpha}{2}}}
    \frac{1}{(\Tilde{u} - u - \bar{u})^{\frac{\beta}{2}} \bar{v}^{\frac{\beta}{2}}}
    \frac{1}{(4uv - 4y^2)^{\frac{\gamma}{2}}} \nn
    = & \MD{(-1)^{\frac{\alpha+\beta+\gamma}{2}}}\frac{a^2 \pi \sqrt{2}}{\bar{v}^{\frac{\alpha + \beta}{2}}} \times \frac{1}{2^{\gamma - 1}} \int du\, d\Tilde{u}\, dv\, dy\, 
    e^{-\frac{1}{2a^2}(u^2 + v^2 + \Tilde{u}^2) - \frac{y^2}{a^2} + i p_u u + i p_v v - 2i p_y y} \nn
    & \quad \times 
    \frac{1}{(\Tilde{u} + u - \bar{u})^{\frac{\alpha}{2}}}
    \frac{1}{(\Tilde{u} - u - \bar{u})^{\frac{\beta}{2}}}
    \frac{1}{u^{\frac{\gamma}{2}} (v - \frac{y^2}{u})^{\frac{\gamma}{2}}}.
\end{align}
At the second equality, the integral with respect to \(\Tilde{v}\) and \(\Tilde{y}\) have been carried out. Next, let us perform the \(v\) integral:
\begin{align}
    \int dv\, e^{-\frac{v^2}{2a^2} + i p_v v} \frac{1}{(v - \frac{y^2}{u})^{\frac{\gamma}{2}}} 
    & = \int dv\, e^{-\frac{1}{2a^2}(v - i p_v a^2)^2 - \frac{a^2 p_v^2}{2}} \frac{1}{(v - \frac{y^2}{u})^{\frac{\gamma}{2}}} \nn
    & = \int dv\, \frac{e^{-\frac{a^2 p_v^2}{2}}}{\Gamma\left(\frac{\gamma}{2}\right)} \frac{1}{v - \frac{y^2}{u}} \frac{\partial^{\frac{\gamma}{2} - 1}}{\partial v^{\frac{\gamma}{2} - 1}} 
    e^{-\frac{1}{2a^2}(v - i p_v a^2)^2} \nn
    & \simeq \frac{2\pi i}{\Gamma\left(\frac{\gamma}{2}\right)} (i p_v)^{\frac{\gamma}{2} - 1} 
    e^{-\frac{y^4}{2a^2 u^2} + i p_v \frac{y^2}{u}}.
\end{align}
Integration by parts is used at the second equality. 
At the third equality, we have kept only the leading term in the limit \eqref{asmp}.

The \( y \) integral is given by  
\begin{align}
        \int dy \, e^{-\frac{y^4}{2a^2u^2}-\frac{y^2}{a^2}+i\frac{p_v}{u} y^2 - 2ip_y y}. 
    \label{yint}
\end{align} 
We evaluate
it
at the stationary phase point \( y=\frac{p_y}{p_v}u \) for large \(p_x, p_y\):  
\begin{align}
    \int dy \, e^{-\frac{y^4}{2a^2u^2}-\frac{y^2}{a^2}+i\frac{p_v}{u}y^2 - 2ip_y y} 
     &\simeq  \, e^{-\frac{u^2}{2a^2} 
    \left( \left(\frac{p_y}{p_v}\right)^4+ 2\left(\frac{p_y}{p_v}\right)^2
    \right) } \int dy \, e^{i\frac{p_v}{u}y^2 - 2ip_y y} \nn
    &=  \, e^{-\frac{u^2}{2a^2} 
    \left( \left(\frac{p_y}{p_v}\right)^4+ 2\left(\frac{p_y}{p_v}\right)^2
    \right) } \sqrt{\frac{ \pi u}{p_v}}\MD{e^{+\frac{\pi}{4}i}}e^{-i\frac{p_y^2}{p_v}u}.
\end{align}
Using this result, the integral of $S$ turns out to be
\begin{align}
\int d^6x S 
     & = \MD{(-1)^{\frac{\alpha+\beta+\gamma}{2}}}\int du\, d\Tilde{u} \,
     i^2\frac{a^2\pi^\frac{5}{2}}{\bar{v}^\frac{\alpha + \beta}{2}2^{\gamma-\frac{5}{2}}}
     \frac{1}{\Gamma(\frac{\gamma}{2})}
     (ip_v)^{\frac{\gamma}{2}-\frac{3}{2}}
     e^{
     -\frac{1}{2a^2}\left(u^2 + \Tilde{u}^2+\left(q^4+2q^2\right)u^2\right)
     \NT{+ 2iku}
     }\nn
     & \quad \times
     \frac{1}{(\Tilde{u}+u-\bar{u})^\frac{\alpha}{2}}
     \frac{1}{(\Tilde{u}-u-\bar{u})^\frac{\beta}{2}}
     \frac{1}{u^{\frac{\gamma}{2}-\frac{1}{2}}}.
     \label{uu}
\end{align}
Here, we introduced
\begin{align}
q \coloneqq  \frac{p_y}{p_v},\qquad
    k \coloneqq \frac{1}{2}\left(p_u - \frac{p_y^2}{p_v}\right)>0.
\end{align}

\PN{By changing the integration variables from \( u,\Tilde{u} \) to \( u_1,u_3 \), we can evaluate the remaining integrals in almost the same manner as in the CFT\(_2\) case. The result is given as (see Appendix~\ref{cal} for details)} 
\begin{align}
    \int d^6x\, A & \simeq C_{TOO}\int d^6x \, S(5, 1, 2\Delta-1) \frac{\bar{v}^2}{4} \simeq 
    \NTmod{(-1)^{2\Delta+1}}
    \frac{4C_{TOO}a^2\pi^\frac{7}{2}}{\PN{3}\bar{v}2^{\Delta-\frac{3}{2}}}
     \frac{p_v^{\Delta-2}}{\Gamma(\Delta-\frac{1}{2})}\frac{k^\Delta}{\Delta \Gamma(\Delta)}e^{-\frac{\bar{u}^2}{2a^2}},\\
     \int d^6 x \, B & \simeq -2C_{TOO}\int d^6x \, S(3, 3, 2\Delta-1) \frac{\bar{v}^2}{4} \simeq
     \NTmod{(-1)^{2\Delta+1}}
     \frac{24C_{TOO}a^2\pi^\frac{7}{2}}{\PN{3}\bar{v}2^{\Delta-\frac{3}{2}}}
     \frac{p_v^{\Delta-2}}{\Gamma(\Delta-\frac{1}{2})}\frac{k^\Delta}{\Delta \Gamma(\Delta)}e^{-\frac{\bar{u}^2}{2a^2}}, \\
     \int d^6x\, C & \simeq C_{TOO}\int d^6x \, S(1, 5, 2\Delta-1) \frac{\bar{v}^2}{4} \simeq
         \NTmod{(-1)^{2\Delta+1}}
     \frac{4C_{TOO}a^2\pi^\frac{7}{2}}{\PN{3}\bar{v}2^{\Delta-\frac{3}{2}}}
     \frac{p_v^{\Delta-2}}{\Gamma(\Delta-\frac{1}{2})}\frac{k^\Delta}{\Delta \Gamma(\Delta)}e^{-\frac{\bar{u}^2}{2a^2}}.
\end{align} 
Since \( D \) is of order \({\cal O}(\bar v^{-3})\) even before integration, it can be neglected in the approximation we have taken.
Thus, we 
find that \eqref{T00_AdS4CFT3} is given by
\begin{align}
     \bra{\omega,p_x,p_y}   T_{00}(t=\bar{t}, x=\bar{x}, y=0) \ket{\omega,p_x,p_y} =& \int d^6x \, (A+B+C+D) \nn
     \simeq & \, 
     \NTmod{(-1)^{2\Delta+1}}
     \frac{32C_{TOO}a^2\pi^\frac{7}{2}}{\PN{3}\bar{v}2^{\Delta-\frac{3}{2}}}
     \frac{p_v^{\Delta-2}}{\Gamma(\Delta-\frac{1}{2})}\frac{k^\Delta}{\Delta \Gamma(\Delta)}e^{-\frac{\bar{u}^2}{2a^2}} \nn
     = &  (-1)^{2\Delta} \frac{4a^2\pi^\frac{5}{2}}{\bar{v}2^{\Delta-\frac{3}{2}}}
     \frac{p_v^{\Delta-2}}{\Gamma(\Delta-\frac{1}{2})}\frac{k^\Delta}{\Gamma(\Delta)}e^{-\frac{\bar{u}^2}{2a^2}},
\end{align}
\NTmod{where \(C_{TOO} = -\frac{3\Delta}{8\pi}\)~\cite{Osborn:1993cr}.}

The normalization factor
\begin{align}
    {\cal{N}}^2 \coloneqq \bra{\omega,p_x,p_y} \omega,p_x,p_y \rangle
\end{align}
can also be computed similarly (see Appendix~\ref{cal} for details):
\begin{align}
    \mathcal{N}^2 
    & = \int d^6x \, S(0, 0, 2\Delta)
    \simeq (-1)^{2\Delta} \frac{
    \NTmod{2}
    a^3\pi^4}{2^{\Delta-\frac{3}{2}}}\frac{(p_v k)^{\Delta-\frac{3}{2}}}{\Gamma\left(\Delta-\frac{1}{2}\right)\Gamma(\Delta)}.
\end{align}
  
Combining the above results, we obtain  
\begin{align}
     {\cal{E}}(\bar{t},\bar{x},0;\omega,p_x,p_y) \coloneqq & \frac{\bra{\omega,p_x,p_y}   T_{00}(t=\bar{t}, x=\bar{x}, y=0) \ket{\omega,p_x,p_y}}{{\cal{N}}^2} \nn
      \simeq & \frac{(\omega^2-p_x^2-p_y^2)^\frac{3}{2}}{\sqrt{2}\pi^{\frac{3}{2}}a(\bar{t}-\bar{x})
      \NTmod{(\omega+p_x)^2}
      }e^{-\frac{(\bar{t}+\bar{x})^2}{2a^2}}.
 \end{align}
Finally, we obtain \( {\cal{E}}(t,x,y;\omega,p_x,p_y) \)
\NTmod{for general $x,y$}
by performing a rotation transformation:  
\begin{align}
 \begin{pmatrix}
     p'_x\\
     p'_y 
 \end{pmatrix}
 = 
 \begin{pmatrix}
     \cos\theta  & -\sin\theta\\
     \sin\theta & \cos\theta
 \end{pmatrix}
 \begin{pmatrix}
     p_x\\
     p_y 
 \end{pmatrix},
 \qquad
 \begin{pmatrix}
     x'\\
     y' 
 \end{pmatrix}
 = 
 \begin{pmatrix}
     \cos\theta  & -\sin\theta\\
     \sin\theta & \cos\theta
 \end{pmatrix}
 \begin{pmatrix}
     \bar{x}\\
     0 
 \end{pmatrix}.
\end{align}
The energy density ${\cal{E}}(t,x',y';\omega,p_x',p_y')$ should 
\NTmod{coincide with ${\cal{E}}(\bar t,\bar x,0;\omega,p_x,p_y)$
by this transformation}
because the energy density is invariant under the spatial rotation.
%
Applying this transformation, we obtain the desired expression as
\begin{align}
     {\cal{E}}(t,x',y';\omega,p_x',p_y')  & = {\cal{E}}(\bar{t},\bar{x},0;\omega,p_x,p_y) \nn
    &\simeq  \frac{(\omega^2 - p_x'^2 - p_y'^2)^{\frac{3}{2}}}{\sqrt{2}\pi^\frac{3}{2}a(t+\sqrt{x'^2+y'^2})\left(\omega - p_x'\frac{x'}{\sqrt{x'^2+y'^2}}-p_y'\frac{y'}{\sqrt{x'^2+y'^2}}\right)^2}e^{-\frac{(t-\sqrt{x'^2+y'^2})^2}{2a^2}}
\end{align}
We omit the primes and denote it simply as ${\cal{E}}(t,x,y;\omega,p_x,p_y)$ hereafter.

To 
confirm energy conservation, we integrate this over space:
\begin{align}
    \int dx\, dy\, {\cal{E}}(t,x,y;\omega,p_x,p_y)
    &=  \int d\theta \, dr \, \frac{r(\omega^2 - \mathbf{p}^2)^{\frac{3}{2}}}{\PN{\sqrt{2}\pi^\frac{3}{2}}a(t+r)\left(\omega - \mathbf{p} \cdot  \mathbf{e}_r \right)^2}e^{-\frac{(t-r)^2}{2a^2}} \nn
    &\simeq  \int d\theta \, \frac{(\omega^2 - \mathbf{p}^2)^{\frac{3}{2}}}{\PN{2\pi}(\omega - p \cos \theta)^2} \nn
    &=   \,\omega, 
\end{align}
where we switched to the polar coordinates.
We obtained the total energy $\omega$, which is consistent with the prepared wave packet state (\ref{wads}).

\section{Discussion}
\label{sec:discussion}

\subsection*{Causality in the AdS$_4$/CFT$_3$ correspondence}

\begin{figure}[htbp]
  \begin{minipage}[t]{0.45\linewidth}
    \centering
    \includegraphics[width=4cm]{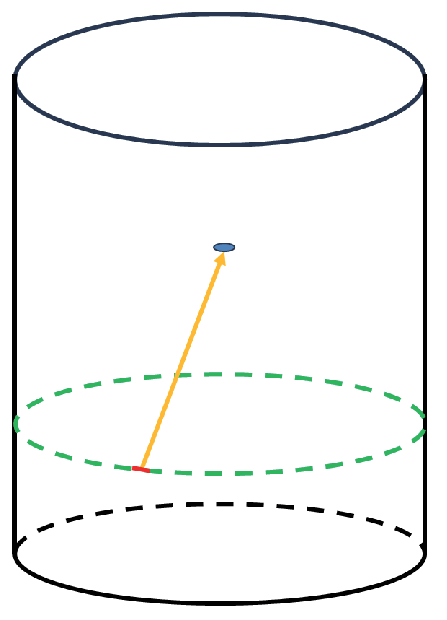}
    \caption{Example of a bulk wave packet moving towards the center.}
    \label{f1}
  \end{minipage}
\hspace{10mm} 
  \begin{minipage}[t]{0.45\linewidth}
    \centering
    \includegraphics[width=4cm]{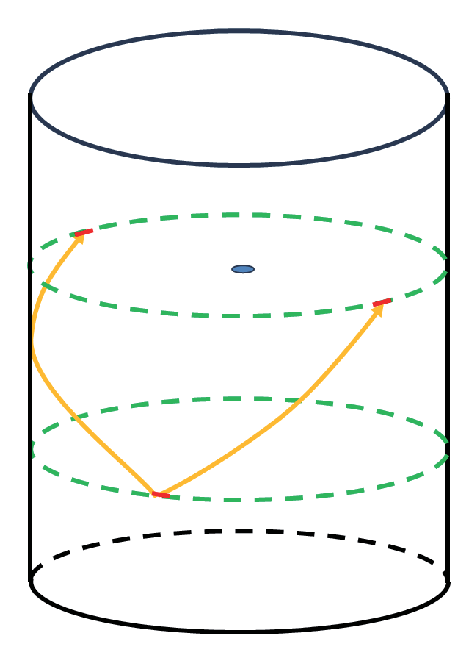}
    \caption{Two ``particles'' on the CFT side corresponding to the bulk wave packet.}
    \label{f2}
  \end{minipage}
\end{figure}

We have considered the wave packets in Poincaré AdS in this work and also in \cite{Terashima:2023mcr}. In the AdS$_3$/CFT$_2$ case discussed in \cite{Terashima:2023mcr}, when switching to the global coordinates, the correspondence depicted by Figures~\ref{f1} and \ref{f2} is obtained. In this case, for a bulk wave packet traveling along a light-like trajectory, two ``particles'' traveling along a light-like path appear as the counterparts in the CFT side. \NT{A bulk wave packet starting from boundary  at \(t=0\) in the bulk reaches the antipodal point on the boundary at \(t=\pi\). At these points, the bulk local field can be identified as the local primary fields at \(t=0\) and \(t=\pi\) on the CFT side, and then the causalities on the both sides consistently hold. Our calculation shows that these properties persist even in the AdS$_4$/CFT$_3$ case; a bulk wave packet traveling at the speed of light corresponds to a localized energy density on the light cone in the CFT side, and causalities on the both sides are preserved without contradiction.}

\subsection*{Wave packet in AdS$_4$/CFT$_3$ correspondence}
In the simplest case, AdS$_3$/CFT$_2$~\cite{Terashima:2023mcr}, it was shown that the energy density on the CFT side \NT{at a given time} is localized at two points on the light cone. This result may be consistent with calculations of excitations by local operators in CFT$_2$~\cite{Caputa:2014vaa, Caputa:2014eta} although these have not studied wave packets.

\begin{figure}
    \centering
    \includegraphics[width=1.02\linewidth]{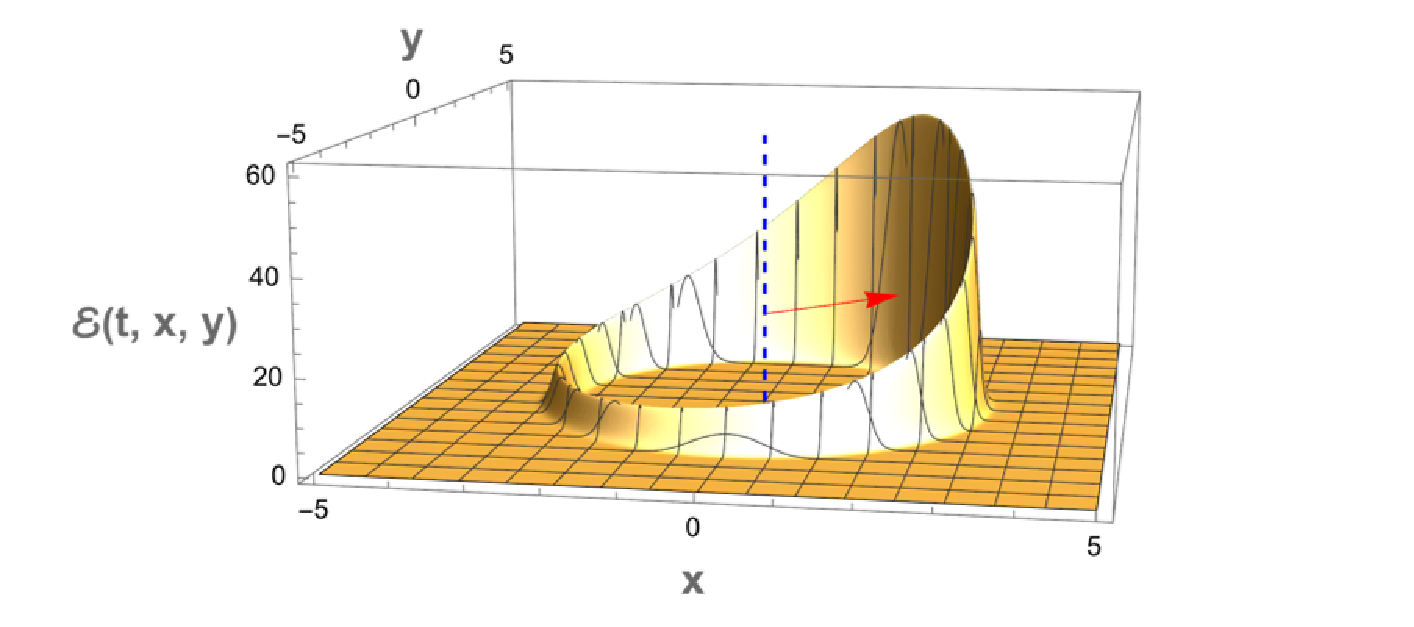}
    \caption{Spatial distribution of the energy density at a certain time \(t\). The red line indicates the direction of the momentum \((p_x,p_y)\).
    The parameters are set as \(t=3,\,a=0.1,\,\omega=15,\,p_x=p_y=5\) in the figure.
    }
    \label{f7}
\end{figure}

Based on our analytical calculation on the CFT side, we found that in AdS$_4$/CFT$_3$, 
the energy density localized on a light cone is excited on the CFT side,
corresponding to a bulk wave packet traveling along a light-like trajectory. This result may also be consistent with the specific limit of holographic energy stress tensor calculations for a heavy particle falling from near the boundary on the bulk side~\cite{Horowitz:1999gf,Nozaki:2013wia}.
Furthermore, in the setup based on~\cite{Terashima:2023mcr}, since the momentum of the bulk wave packet serves as a parameter, we were able to observe the correspondence between the direction of momentum and the energy distribution. The result is
\begin{align}
     {\cal{E}}(t,r,\theta;\omega,\mathbf{p}) 
    \simeq \, \frac{(\omega^2 - \mathbf{p}^2)^{\frac{3}{2}}}{\sqrt{2}\pi^\frac{3}{2}a(t+r)\left(\omega - \mathbf{p} \cdot  \mathbf{e}_r \right)^2}e^{-\frac{(t-r)^2}{2a^2}}.
    \label{energy-density}
\end{align}
The exponential factor indicates
that the energy density is localized on the light cone, while the \MD{denominator shows that} the distribution concentrates along the momentum direction
as illustrated in Figure~\ref{f7}.

The dependence of the energy density \eqref{energy-density} on $\omega$ and $\mathbf{p}$ is naturally understood as the outcome of the relativistic
Doppler effect
due to the motion of the wave source. Consider a circularly symmetric thin shell of light-like particles whose energy density is isotropic. The energy-momentum tensor at the shell surface is given by 
\NTmod{$T^{\mu\nu} = \mathcal{E}'  n^\mu n^\nu$,}
where $n^\mu = (1,\mathbf{e}'_r ) $ is tangent to the propagation direction and $\rho'$ is the surface energy density that is constant on the shell surface. Introduce a boosted observer whose 3-velocity is $u^\mu = \gamma (1, -\mathbf{v})$, 
where $\gamma=(1-|\mathbf{v}|)^{-1/2}$ is the gamma factor.
The energy density for this observer is given by $\mathcal{E} = T_{\mu\nu} u^\mu u^\nu = \mathcal{E}' \left[\gamma (1+\mathbf{v}\cdot\mathbf{e}'_r )\right]^2 = \mathcal{E}' \left[\gamma (1+|\mathbf{v}| \cos\theta' )\right]^2$, where and $\theta'$ is the angle between $\mathbf{v}$ and $\mathbf{e}'_r$ measured in the rest frame of the source of the light-like particles. Transforming the angle $\theta'$ to that in the observer rest frame $\theta$ by the formula for the relativistic beaming effect $\cos\theta' = (\cos\theta - |\mathbf{v}|)/(1-|\mathbf{v}|\cos\theta)$, we find
\begin{equation}
\mathcal{E} = \frac{\mathcal{E}'}{\gamma^2(1-|\mathbf{v}|\cos\theta)^2}.
\label{rho-rho'}
\end{equation}
This is the energy density distribution on the circularly symmetric shell of the light-like particles.
Based on this observation, let us express
\eqref{energy-density} in the $a\to 0$ limit as
\begin{align}
     {\cal{E}}(t,r,\theta;\omega,\mathbf{p}) 
    \sim 
    \frac{1}{
    \gamma^2(1-|\mathbf{v}|\cos\theta)^2}
    \NTmod{
    \frac{\sqrt{\omega^2-|\mathbf{p}|^2}}{\pi}
    \delta\bigl(t^2-r^2\bigr)},
    \label{energy-density2}
\end{align}
which share the same form as \eqref{rho-rho'}.
$\mathbf{v} \coloneqq \mathbf{p}/\omega$ 
is the velocity of the frame in which the energy density on the light cone becomes isotropic.
\NTmod{$\frac{\sqrt{\omega^2-|\mathbf{p}|^2}}{\pi}
    \delta\bigl(t^2-r^2\bigr) = \frac{\gamma^{-1}\omega}{2\pi r}\delta(t-r)$ 
corresponds to $\mathcal{E}'$ introduced above, where $\gamma^{-1}\omega$ is the total energy measured in the frame with velocity $\mathbf{v}$.}

The boost considered in the above discussions is easily understood in the bulk picture
because the Lorentz transformations for the $d$-dimensional Minkowski spacetime can be trivially lifted to the Poincare patch of the AdS$_{d+1}$ spacetime.
First, we consider a wave packet moving along the radial direction, which has $\mathbf{p}=0$.
The corresponding energy 
\NTmod{distribution in the CFT}
should be a circular one.
The boost with the velocity $\mathbf{v}=\mathbf{p}/\omega$ in the bulk makes the wave packet
with $(\omega,\mathbf{p})$, where 
we had chosen the frequency of the original wave packet appropriately.
Then, the energy density in the CFT will be given by the boosted one, i.e. \eqref{energy-density2} in the $a \rightarrow 0$ limit.

In previous studies for AdS\(_3\)/CFT\(_2\), a particle-like description existed on the CFT side corresponding to a bulk wave packet. However, in the case of AdS\(_4\)/CFT\(_3\), energy density spreads out spatially at a given time, and a particle-like description does not apply.

\subsection*{Weaker version of the entanglement wedge reconstruction}
Our results are consistent with the weaker version of the entanglement wedge reconstruction~\cite{Terashima:2020uqu, Sugishita:2022ldv, Sugishita:2023wjm, Sugishita:2024lee}. This version claims the following:
a bulk operator with \(1/N\) corrections in the entanglement wedge corresponding to a CFT subregion can be reconstructed from the operator in the CFT subregion, whereas bulk operators outside the entanglement wedge cannot be reconstructed.\footnote{Some argue that EWR holds even with 
$1/N$ corrections in the sense of~\cite{Dong:2016eik}, while others, such as~\cite{Cotler:2017erl}, take the position that introducing $1/N$ corrections causes it to break down. 
The weaker version of the entanglement wedge reconstruction is different from those.
}
When considering \(1/N\) corrections in the bulk, gravitational dressing is required. In this context, the region on the CFT side where energy is nonzero can be considered as the region where the gravitational dressing has endpoints. By selecting a subregion on the CFT side that includes the region with nonzero energy, the bulk operator may be reconstructed. In this sense, the results of this study are consistent with the weaker version of the entanglement wedge reconstruction.

\section*{Acknowledgement}

The authors would like to thank S.~Sugishita and T.~Takayanagi for their helpful discussions.
This work was supported by MEXT-JSPS Grant-in-Aid for Transformative Research Areas~(A) ``Extreme Universe'', Nos.~21H05184 and 21H05189.
This work was supported in part by JSPS KAKENHI Grant Numbers~22H05111 and 24K07048.


\appendix
\section{Details of calculation}\label{cal}
In this appendix, we show the detailed calculations for the terms \(A, B, C\) and normalization \(\mathcal{N}^2\) introduced in section~\ref{sec:AdS4CFT3}.
Using \(u=\frac{1}{2}(u_1-u_3),\, \Tilde{u}=\frac{1}{2}(u_1+u_3)\),
\eqref{uu} is given in terms of $u_1$ and $u_3$ as
\begin{align}
     \int d^6x S(\alpha,\beta, \gamma)
     & =\MD{(-1)^{\frac{\alpha+\beta+\gamma}{2}}}\int\, du_1\,du_3\, i^2\frac{a^2\pi^\frac{5}{2}}{\bar{v}^\frac{\alpha + \beta}{2}2^{\frac{\gamma}{2}-1}}
     \frac{1}{\Gamma(\frac{\gamma}{2})}
     (ip_v)^{\frac{\gamma}{2}-\frac{3}{2}}
     e^{
     -\frac{1+R}{4a^2}\left(u_1^2+u_3^2-\frac{2R}{1+R}u_1u_3\right)+iku_1-iku_3
     } \nn
     & \quad \times \frac{1}{(u_1-\bar{u})^\frac{\alpha}{2}} \frac{1}{(u_3-\bar{u})^\frac{\beta}{2}}\frac{1}{(u_1-u_3)^{\frac{\gamma}{2}-\frac{1}{2}}},
     \label{general}
\end{align}
where we defined \(R\coloneqq \frac{q^4+2q^2}{2}\) using \(q=\frac{p_y}{p_v}\).
%
%
\(\gamma\) is common to \(A, B, C\), and it is given by \(\gamma=2\Delta-1\). \(\alpha + \beta\) and \(\alpha + \beta + \gamma\) are also common, being equal to \(6\) and \(2\Delta + 5\), respectively.
We differentiate \( \Delta - 1 \) times with respect to \( k \) before performing this integral. By performing the differentiation with respect to \(k\), these calculations become the same as in the CFT\(_2\) case 
for $\Delta \in \mathbf{Z}$. For $\Delta \notin \mathbf{Z}$, we can compute it in the same way by differentiating \( \lfloor\Delta - 1\rfloor \) times, where $\lfloor x\rfloor $ is the function that takes as input a real number $x$, and gives as output the greatest integer less than or equal to $x$, i.e. the floor function. For simplicity, we will concentrate on the $\Delta \in \mathbf{Z}$ case below because the result is correct even for $\Delta \notin \mathbf{Z}$.
\begin{align}
     \int d^6x S(\alpha,\beta, \gamma)
     & = \MD{(-1)^{\PN{\Delta}}}\PN{i^{\Delta-1}}\PN{\left(\int dk\right)^{\Delta-1}}\int\, du_1\,du_3\, \PN{i^3}\frac{a^2\pi^\frac{5}{2}}{\bar{v}^32^{\Delta-\frac{3}{2}}}
     \frac{1}{\Gamma(\Delta-\frac{1}{2})}
     (ip_v)^{\Delta-2}
     \nn
     & \quad \times e^{
     -\frac{1+R}{4a^2}\left(u_1^2+u_3^2-\frac{2R}{1+R}u_1u_3\right)+iku_1-iku_3
     } \frac{1}{(u_1-\bar{u})^\frac{\alpha}{2}} \frac{1}{(u_3-\bar{u})^\frac{\beta}{2}} \nn
     & \simeq \MD{(-1)^{\Delta}}\PN{i^{\Delta-1}}\PN{\left(\int dk\right)^{\Delta-1}}\int\, du_3\, \PN{i^3}\frac{a^2\pi^\frac{5}{2}}{\bar{v}^32^{\Delta-\frac{3}{2}}}
     \frac{1}{\Gamma(\Delta-\frac{1}{2})}
     (ip_v)^{\Delta-2}
     \nn
     & \quad \times e^{
     -\frac{1+R}{4a^2}\left(\bar{u}^2+u_3^2-\frac{2R}{1+R}\bar{u}u_3\right)+ik\bar{u}-iku_3
     } \frac{2\pi i}{\Gamma\left(\frac{\alpha}{2}\right)}(ik)^{\frac{\alpha}{2}-1}\frac{1}{(u_3-\bar{u})^\frac{\beta}{2}}\nn
     & \simeq \MD{(-1)^{\Delta}}\PN{i^{\Delta-1}}\PN{\left(\int dk\right)^{\Delta-1}} \PN{i^3}\frac{a^2\pi^\frac{5}{2}}{\bar{v}^32^{\Delta-\frac{3}{2}}}
     \frac{1}{\Gamma(\Delta-\frac{1}{2})}
     (ip_v)^{\Delta-2}
     \nn
     & \quad \times e^{
     -\frac{\bar{u}^2}{2a^2}
     } \frac{2\pi i}{\Gamma\left(\frac{\alpha}{2}\right)}(ik)^{\frac{\alpha}{2}-1}\frac{-2\pi i}{\Gamma\left(\frac{\beta}{2}\right)}(-ik)^{\frac{\beta}{2}-1} \nn
     & = (-1)^P(-1)^{2\Delta} \frac{4a^2\pi^\frac{9}{2}}{\bar{v}^32^{\Delta-\frac{3}{2}}}
     \frac{p_v^{\Delta-2}}{\Gamma(\Delta-\frac{1}{2})}\frac{k^\Delta}{\Delta \Gamma(\Delta)}e^{-\frac{\bar{u}^2}{2a^2}}\frac{1}{\Gamma\left(\frac{\alpha}{2}\right)\Gamma\left(\frac{\beta}{2}\right)},
\end{align}
\NTmod{where \(P\) is odd for \(A\) and \(C\), but even for \(B\).}
At the second equality, we perform \(u_1\) integral. At the third equality, we perform \(u_3\) integral, where the integration contour lies in the lower half-plane, which introduces a minus sign. 
\PN{
%
In the above result, there is a subtlety in the overall sign; that is, the overall sign may change if we have chosen different branches in evaluating numerical factors, such as $(-1)^{\frac{\alpha+\beta+\gamma}{2}}$ with non-integer power. This ambiguity can be fixed by carefully treating the analytic continuation from the Euclidean formulation. In this work, however, we choose it to ensure the positivity of the final result for simplicity.
}

Using this expression, we can easily evaluate the terms \(A, B, C\).
\begin{align}
    \int d^6x\, A & \simeq C_{TOO}\int d^6x \, S(5, 1, 2\Delta-1) \frac{\bar{v}^2}{4} \simeq 
    \NTmod{(-1)^{2\Delta+1}}
    \frac{4C_{TOO}a^2\pi^\frac{7}{2}}{\PN{3}\bar{v}2^{\Delta-\frac{3}{2}}}
     \frac{p_v^{\Delta-2}}{\Gamma(\Delta-\frac{1}{2})}\frac{k^\Delta}{\Delta \Gamma(\Delta)}e^{-\frac{\bar{u}^2}{2a^2}},\nn
     \int d^6 x \, B & \simeq -2C_{TOO}\int d^6x \, S(3, 3, 2\Delta-1) \frac{\bar{v}^2}{4} \simeq
     \NTmod{(-1)^{2\Delta+1}}
     \frac{24C_{TOO}a^2\pi^\frac{7}{2}}{\PN{3}\bar{v}2^{\Delta-\frac{3}{2}}}
     \frac{p_v^{\Delta-2}}{\Gamma(\Delta-\frac{1}{2})}\frac{k^\Delta}{\Delta \Gamma(\Delta)}e^{-\frac{\bar{u}^2}{2a^2}}, \nn
     \int d^6x\, C & \simeq C_{TOO}\int d^6x \, S(1, 5, 2\Delta-1) \frac{\bar{v}^2}{4} \simeq
         \NTmod{(-1)^{2\Delta+1}}
     \frac{4C_{TOO}a^2\pi^\frac{7}{2}}{\PN{3}\bar{v}2^{\Delta-\frac{3}{2}}}
     \frac{p_v^{\Delta-2}}{\Gamma(\Delta-\frac{1}{2})}\frac{k^\Delta}{\Delta \Gamma(\Delta)}e^{-\frac{\bar{u}^2}{2a^2}}.
\end{align} 
Therefore 
\begin{align}
     \bra{\omega,p_x,p_y}   T_{00}(t=\bar{t}, x=\bar{x}, y=0) \ket{\omega,p_x,p_y} =& \int d^6x \, (A+B+C+D) \nn
     \simeq & \,
     \NTmod{(-1)^{2\Delta+1}}
     \frac{32C_{TOO}a^2\pi^\frac{7}{2}}{\PN{3}\bar{v}2^{\Delta-\frac{3}{2}}}
     \frac{p_v^{\Delta-2}}{\Gamma(\Delta-\frac{1}{2})}\frac{k^\Delta}{\Delta \Gamma(\Delta)}e^{-\frac{\bar{u}^2}{2a^2}} \nn
     = &  (-1)^{2\Delta} \frac{4a^2\pi^\frac{5}{2}}{\bar{v}2^{\Delta-\frac{3}{2}}}
     \frac{p_v^{\Delta-2}}{\Gamma(\Delta-\frac{1}{2})}\frac{k^\Delta}{\Gamma(\Delta)}e^{-\frac{\bar{u}^2}{2a^2}},
\end{align}
\NTmod{where \(C_{TOO} = -\frac{3\Delta}{8\pi}\)~\cite{Osborn:1993cr}.}

We can easily evaluate the normalization factor \(\mathcal{N}^2\) by using \eqref{uu}.
It suffices to set \(\alpha=\beta=0,\) and \(\gamma = 2\Delta\). 
\begin{align}
    \mathcal{N}^2 
    & = \int d^6x \, S(0, 0, 2\Delta) \nn
    & = (-1)^\Delta 2 \int du\, d\Tilde{u}\, i^2 \frac{a^2 \pi^\frac{5}{2}}{
    \NTmod{2^{2\Delta-\frac{3}{2}}}
    } \frac{(ip_v)^{\Delta-\frac{3}{2}}}{\Gamma(\Delta)}e^{-\frac{\Tilde{u}^2}{2a^2}}e^{-\frac{(1+2R)u^2}{2a^2}+2iku} \frac{1}{u^{\Delta-\frac{1}{2}}} \nn
    & = (-1)^\Delta 2\sqrt{2} \int du\, i^2 \frac{a^3 \pi^3}{
    \NTmod{2^{2\Delta-\frac{3}{2}}}
    } \frac{(ip_v)^{\Delta-\frac{3}{2}}}{\Gamma(\Delta)}e^{-\frac{(1+2R)u^2}{2a^2}+2iku} \frac{1}{u^{\Delta-\frac{1}{2}}} \nn 
    & \simeq (-1)^\Delta i^2\frac{2\sqrt{2}a^3 \pi^3}{
    \NTmod{2^{2\Delta-\frac{3}{2}}}
    }\frac{(ip_v)^{\Delta-\frac{3}{2}}}{\Gamma(\Delta)}\frac{2\pi i
    (2ik)^{\Delta-\frac{3}{2}}
    }{\Gamma\left(\Delta-\frac{1}{2}\right)} \nn
    & = (-1)^{2\Delta} \frac{
    \NTmod{2}
    a^3\pi^4
    }{2^{\Delta-\frac{3}{2}}}\frac{(p_v k)^{\Delta-\frac{3}{2}}}{\Gamma\left(\Delta-\frac{1}{2}\right)\Gamma(\Delta)}
\end{align}
Combining the above results, we obtain  
\begin{align}
     {\cal{E}}(\bar{t},\bar{x},0;\omega,p_x,p_y) \coloneqq & \frac{\bra{\omega,p_x,p_y}   T_{00}(t=\bar{t}, x=\bar{x}, y=0) \ket{\omega,p_x,p_y}}{{\cal{N}}^2} \nn
      \simeq & \frac{
      \NTmod{2}
      (p_vk)^\Delta}{
      \pi^\frac{3}{2}a\bar{v} p_v^2 }   e^{-\frac{\bar{u}^2}{2a^2}} \nn
       = & \frac{(\omega^2-p_x^2-p_y^2)^\frac{3}{2}}{\sqrt{2}\pi^{\frac{3}{2}}a(\bar{t}-\bar{x})
       \NTmod{(\omega+p_x)^2}
       }e^{-\frac{(\bar{t}+\bar{x})^2}{2a^2}}.
\end{align}

\bibliographystyle{utphys}

\input{english_main.bbl}

\end{document}

%% file: english_main.bbl
\providecommand{\href}[2]{#2}\begingroup\raggedright\endgroup